\documentclass[nolinenumbers, twocolumn]{aastex63}
\usepackage[utf8]{inputenc}

\def\kms{km$\ $s$^{-1}$}
\def\lamost{J0240}

\begin{document}

\title{Confirmation of a Second Propeller: A High-Inclination Twin of AE~Aquarii}

\author{Peter Garnavich}
\affiliation{University of Notre Dame, Notre Dame, IN 46556, USA}
\author{Colin Littlefield}
\affiliation{University of Notre Dame, Notre Dame, IN 46556, USA}
\affiliation{Department of Astronomy, University of Washington, Seattle, WA 98195, USA}
\author{R. M. Wagner}
\affiliation{Department of Astronomy, Ohio State University, Columbus, OH 43210, USA}
\author{Jan van Roestel}
\affiliation{Departments of Astronomy and Physics, California Institute of Technology, Pasadena CA 91125, USA}
\author{Amruta D. Jaodand}
\affiliation{Departments of Astronomy and Physics, California Institute of Technology, Pasadena CA 91125, USA}
\author{Paula Szkody}
\affiliation{Department of Astronomy, University of Washington, Seattle, WA 98195, USA}
\author{John R. Thorstensen}
\affiliation{Department of Physics and Astronomy, Dartmouth College, Hanover, NH 03755 USA}

\begin{abstract}

    For decades, AE~Aquarii (AE~Aqr) has been the only cataclysmic variable star known to contain a magnetic propeller: a persistent outflow whose expulsion from the binary is powered by the spin-down of the rapidly rotating, magnetized white dwarf. In 2020, LAMOST~J024048.51+195226.9 (\lamost) was identified as a candidate eclipsing AE~Aqr object, and we present three epochs of time-series spectroscopy that strongly support this hypothesis. We show that during the photometric flares noted by \citet{thorstensen20}, the Balmer and He~I emission lines reach velocities of $\sim$3000~\kms, well in excess of what is observed in normal cataclysmic variables. This is, however, consistent with the high-velocity emission seen in flares from AE~Aqr. Additionally, we confirm beyond doubt that \lamost\ is a deeply eclipsing system. The flaring continuum,  He~I and much of the Balmer emission likely originate close to the WD because they disappear during the eclipse that is centered on inferior conjunction of the secondary star. The fraction of the Balmer emission remaining visible during eclipse is likely produced in the extended outflow. Most enticingly of all, this outflow produces a narrow P~Cygni absorption component for nearly half of the orbit, and we demonstrate that this scenario closely matches the outflow kinematics predicted by \citet{wynn97}. While an important piece of evidence for the magnetic-propeller hypothesis---a rapid WD spin period---remains elusive, our spectra provide compelling support for the existence of a propeller-driven outflow viewed nearly edge-on, enabling a new means of rigorously testing theories of the propeller phenomenon.

\end{abstract}


\section{Introduction}

Intermediate polars (IPs, aka DQ~Herculis stars) are a type of cataclysmic variable star (CV) consisting of a cool, non-degenerate companion transferring mass to a magnetized white dwarf (WD) whose spin period is significantly shorter than the binary orbit \citep{patterson94}.  IPs generally accrete from a truncated accretion disk whose inner rim extends to the edge of the WD's magnetosphere. Once the magnetic pressure exceeds the disk's ram pressure, the gas begins to travel along the WD's magnetic fields lines. The magnetically confined part of the flow is often called an ``accretion curtain,'' and as the WD rotates, the changing aspect of the curtain causes a photometric modulation at the WD's spin period \citep{patterson94}. However, in some IPs, the WD's magnetic field is large enough to prevent the formation of an accretion disk, and in these diskless systems, the ballistic accretion stream from the companion star directly impacts the WD's magnetosphere \citep[e.g., V2400~Oph:][]{v2400_oph}.

If the magnetized WD is spinning sufficiently fast, donated gas may be ejected from the system via a ``magnetic propeller'' mode \citep{wynn97}, in which the rapidly rotating magnetosphere of the WD acts as a centrifugal barrier that inhibits accretion onto the WD. The magnetic propeller mechanism is believed to operate over a wide range of accreting systems, including neutron stars and young stellar objects \citep{propellers}, but in WD systems, the propeller mode appears extremely rare.

Until now, the only confirmed IP in a propeller mode has been AE~Aqr \citep{eracleous96, wynn97}, which displays unique photometric and spectroscopic properties, including a WD spin period of 33~s \citep{patterson79}. \citet{meintjes} review the major observational and theoretical studies of AE~Aqr. Several additional IPs with WD rotation periods near 30~s have recently been identified \citep[e.g.][]{oliviera20, ashley20}, but their magnetic fields are apparently insufficient to power an AE~Aqr-like propeller. AE~Aqr and these other IPs may require a recent episode of thermal time-scale mass transfer to spin-up their WDs to these extreme rates \citep{schenker02}.

Recently, \citet{thorstensen20} pointed to the flaring light curves and spectral properties of LAMOST~J024048.51+195226.9 (\lamost\ hereafter) as being similar to those of AE~Aqr and suggested that it might be only the second IP in the propeller mode. \citet{littlefield20} analyzed its long-term light curve from sky surveys and concluded that the secondary star eclipses the location of the flaring activity, implying a high orbital inclination. The system's relatively long orbital period of 7.33~h \citep{drake14,thorstensen20} has been refined by \citet{littlefield20} to a precise value of 0.3056849(5)~d.  
Here, we present fast-cadence, time-resolved spectroscopy and photometry to compare \lamost\ with the observational properties of original propeller AE~Aqr.

\begin{figure}
    \centering
    \includegraphics[width=\columnwidth]{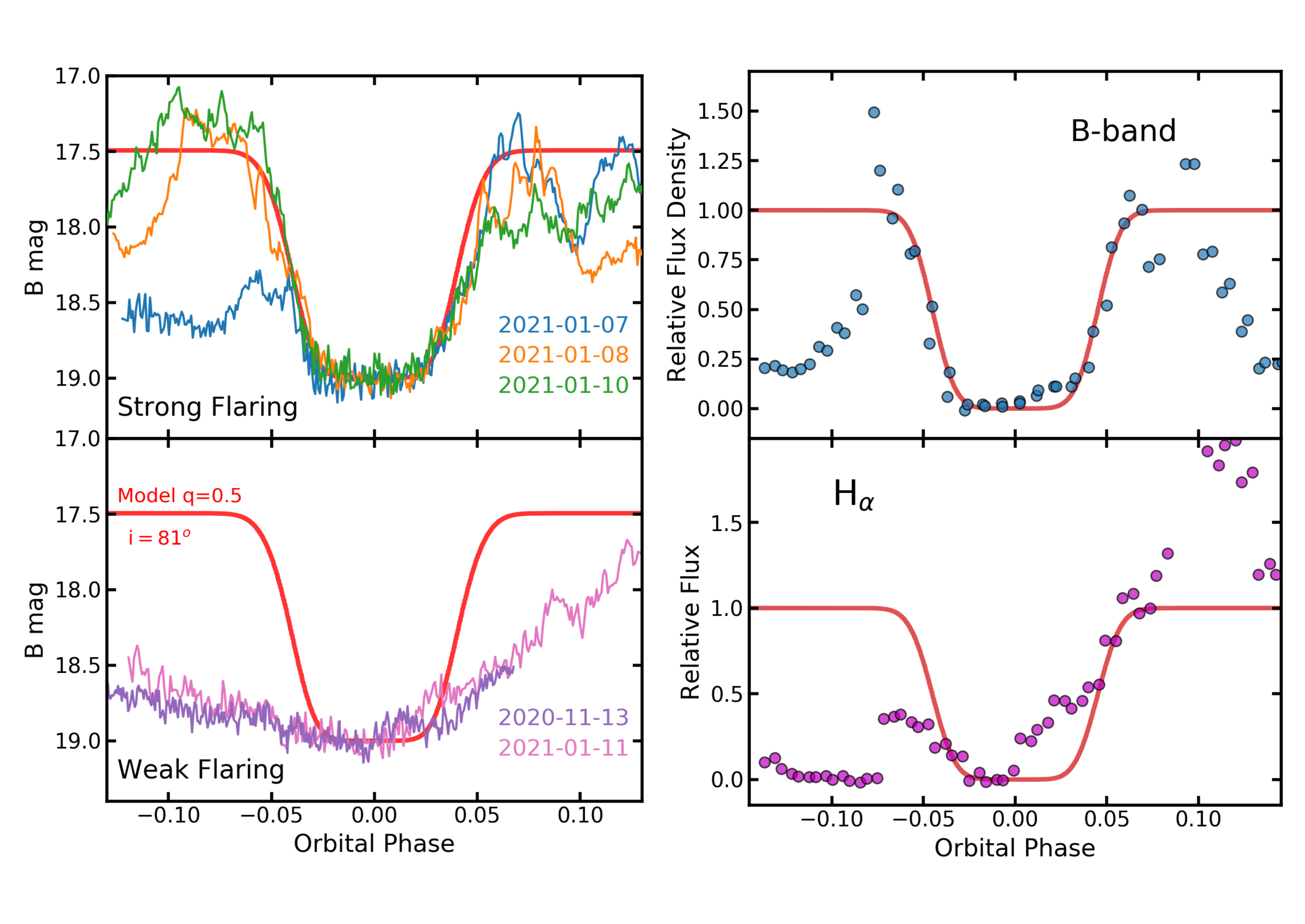}
    \caption{ Light curves around the time of inferior conjunction. During periods of active flaring the outline of the eclipse is clearly seen. The red lines are a model eclipse of a constant source located at the position of a WD as described in the text. {\bf left:} Five nights of MDM $B$-band photometry. The three runs where flaring was active is plotted in the top panel, while two quiescent eclipses are shown in the lower panel.   {\bf right:} The blue flux density and H$_\alpha$ emission line flux from the September $LBT$ spectra are displayed with an eclipse model. A small rise in the continuum flux accompanied by a larger brightening in the emission line flux begins precisely at inferior conjunction.   }
    \label{eclipse_lc}
\end{figure}

\begin{figure*}[t]
    \centering
    \includegraphics[width=\textwidth]{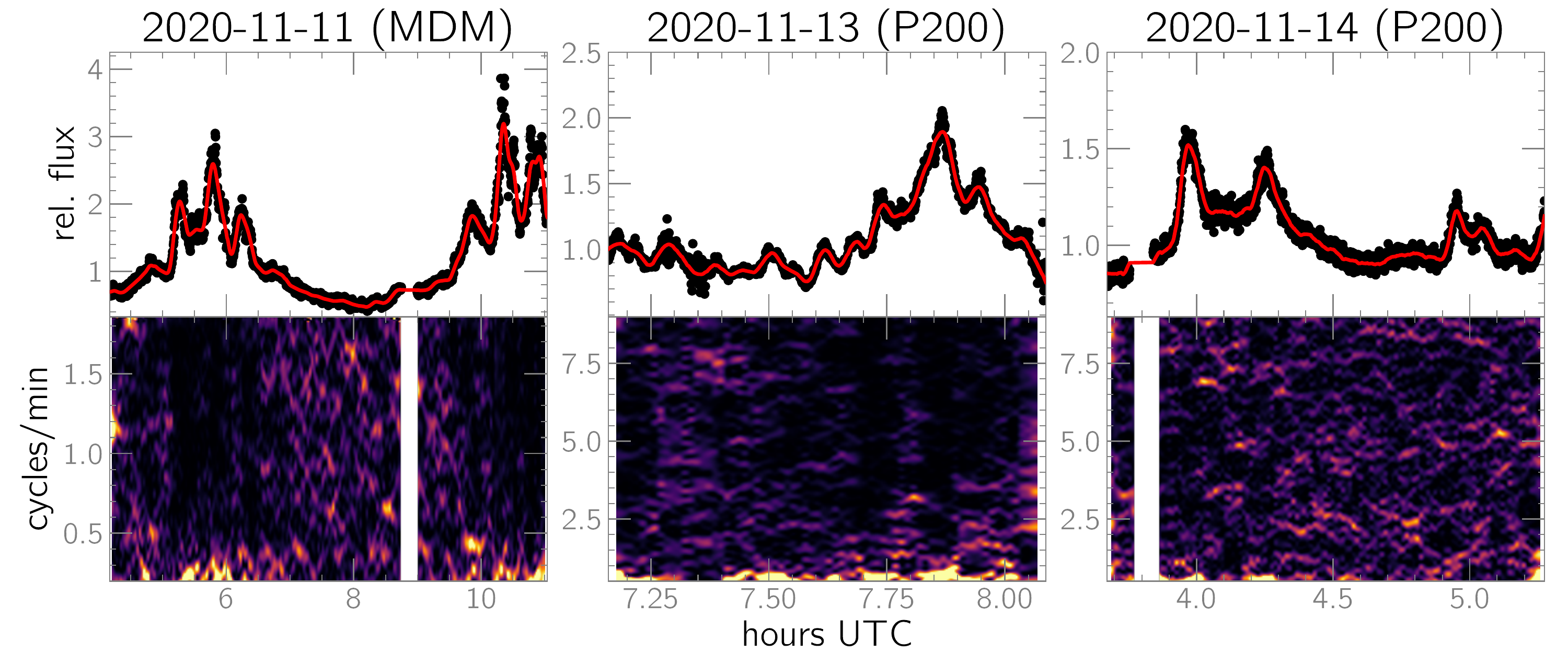}
    \caption{The MDM and Palomar light curves, with their two-dimensional power spectra. The smooth red curves were subtracted from the observed data (black points) prior to computation of the power spectra. There is no convincing evidence of a stable period in any of these light curves.}
    \label{fig:trailed_power}
\end{figure*}

\begin{figure*}[t]
\centering
\includegraphics[width=\textwidth]{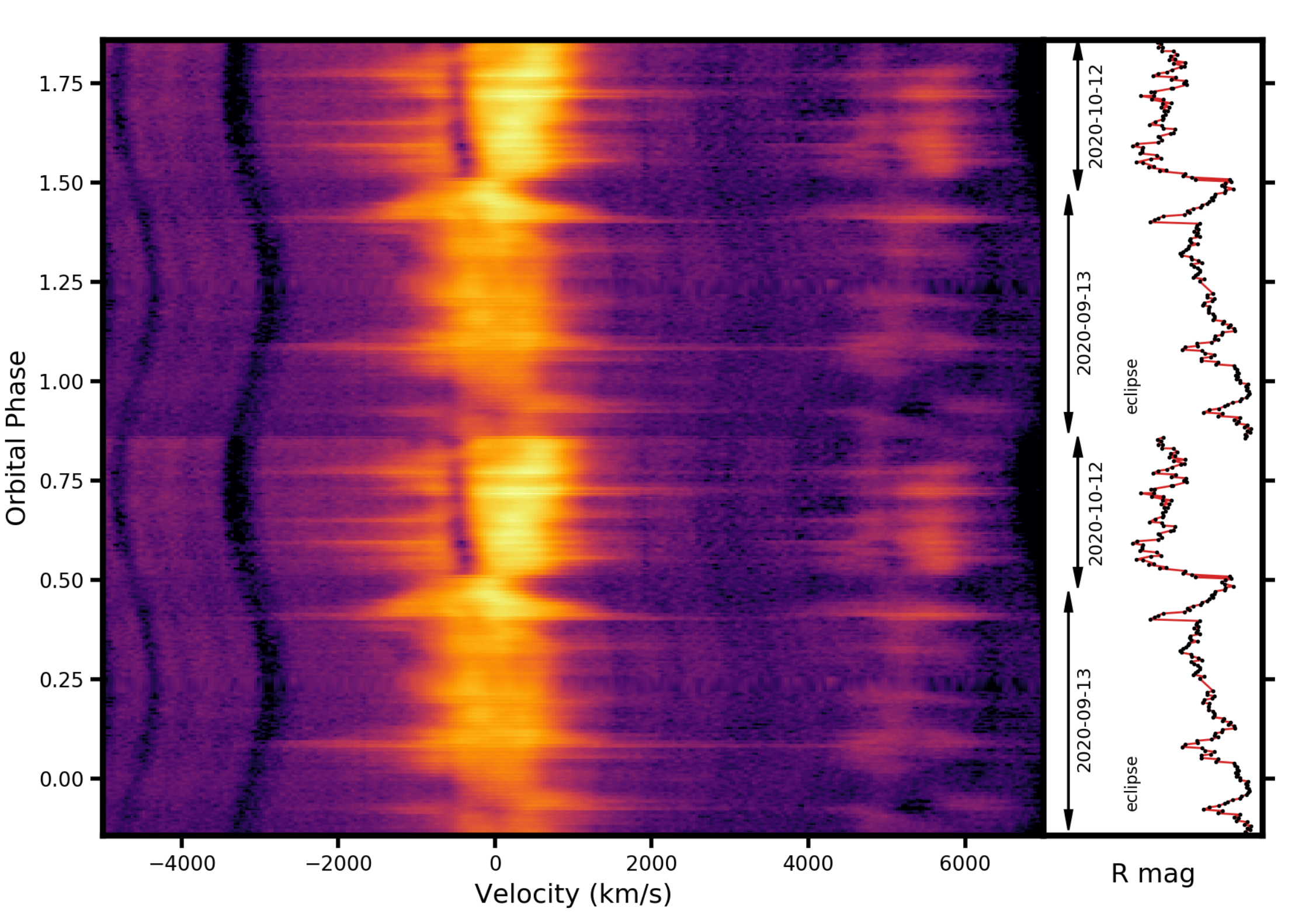}
\caption{{\bf left:} Trailed spectra of \lamost\ around H$_\alpha$ and He~I 6675~\AA . A full binary orbit is shown and the sequence is repeated. Several absorption lines from the secondary are visible in the continuum. The full orbit is constructed by combining the September and October runs resulting in discontinuities in the emission profiles at phases 0.50 and 0.84. {\bf right:} The $R$-band light curve synthesized from the individual spectra. The star was brighter and flaring more active during the October run.  }
\label{stack}
\end{figure*}

\begin{deluxetable*}{lccccc}
\centering
\tablecaption{Photometric Observations \label{photometry}}
\tablehead{
\colhead{Date} & \colhead{Start Time} & \colhead{Length} & \colhead{Cadence\tablenotemark{\tiny a}} & \colhead{Filter} & \colhead{Telescope} \vspace{-0.2cm}
\\
\colhead{(UT)} & \colhead{(UT)} & \colhead{(hours)} & \colhead{(s)} & & 
}
\startdata
2020-11-11 & 04:01  & 6.87 & 16.5 & $B$ & MDM \\
2020-11-13 & 07:09  & 0.93 & 3.0 & $g'$,$i'$ & P200 \\
2020-11-14 & 03:40  & 1.60 & 3.0 & $g'$,$i'$ & P200 \\
2021-01-07 & 03:51  & 1.96 & 20.4 & $B$ & MDM \\
2021-01-08 & 01:50  & 2.00 & 20.4 & $B$ & MDM \\
2021-01-10 & 05:03  & 2.15 & 20.4 & $B$ & MDM \\
2021-01-11 & 03:15  & 2.02 & 20.4 & $B$ & MDM \\
\enddata
\tablenotetext{\tiny a}{Cadence is the average time between consecutive exposures.}
\end{deluxetable*}

\section{Data}

\subsection{Spectroscopy}

We observed \lamost\ with the Large Binocular Telescope (LBT\footnote{The LBT is an international collaboration among institutions in the United States, Italy and Germany. LBT Corporation partners are: The University of Arizona on behalf of the Arizona Board of Regents; Istituto Nazionale di Astrofisica, Italy; LBT Beteiligungsgesellschaft, Germany, representing the Max-Planck Society, The Leibniz Institute for Astrophysics Potsdam, and Heidelberg University; The Ohio State University, representing OSU, University of Notre Dame, University of Minnesota and University of Virginia.}) and twin Multi-Object Dual Spectrographs \citep[MODS;][]{pogge12} on 2020 September 12, October 13, and November 20 (UT). The spectrographs were set up in dual grating mode providing wavelength coverage from 320~nm to 1.01~$\mu$m, with a dichroic splitting the light into the red and blue optimized arms at 565~nm. A 0.8~arcsec slit was employed on all the nights giving a spectral resolution of R=1350 near H$_\alpha$. The typical seeing during the observations was between 0.8 and 1.0~arcsec. The sky was clear and seeing was steady around 1.0~arcsec for the September and October runs. In November, there were some clouds and the seeing varied between 1.2 and 1.5~arcsec.

The four MODS spectrographs (SX/MODS1, DX/MODS2 plus the red and blue arms for both) were run independently and each have slightly different readout and overhead costs. So, despite all the spectrographs being set to take 180s integrations, the start times for a long series of exposures soon become unsynchronized. This results in an improved temporal resolution when data from the two telescopes are combined. On the September night, 135 red exposures were obtained covering 4.7 hours while on the October, 72 red-side exposures were taken covering 2.5~hours. The average cadence was 2~minutes for the red spectra. Serendipitiously, the September run covered orbital phases 0.87-1.51, while the October data spanned orbital phases 0.52-0.87, so the two runs combined covered an entire 0.306~day binary orbit. The November run generated 75 red-side spectra that covered orbital phases 0.61-0.98, significantly overlapping with phases observed in October.

The two-dimensional CCD images were processed and one-dimensional spectra extracted using a 1.0~arcsec aperture. The spectra were wavelength calibrated using neon, argon, and mercury emission lamps. To correct for flexture over the time series, the wavelength scale was adjusted slightly to shift the airglow emission lines to their rest wavelengths. The \lamost\ spectra were flux calibrated using the spectrophotometric standard star BD+28$^\circ$4211. Synthetic $B$, $V$, and $R$ magnitudes were estimated by averaging the flux densities over the Johnson-Cousins bandpasses. 

\subsection{Photometry}

We obtained time-resolved photometry with the MDM 2.4m Hiltner telescope and OSMOS camera on five nights in late 2020 and early 2021. A log of the observations is given in Table~\ref{photometry}. The first run covered nearly an entire binary orbit with an average time between exposures (the cadence) of 16.5~s. The observations in 2021 January were timed to cover two hours centered around inferior conjunction with a slightly longer cadence of 20.4~s. The data were obtained through a Johnson $B$ filter and calibrated assuming that a comparison star 91$''$E and 25$''$N of \lamost\ has an apparent brightness of $B=16.00\pm0.06$ mag\footnote{From the APASS catalog \citep{APASS}}. 

Very high cadence photometry was also obtained with the Palomar 200-inch telescope (P200) and CHIMERA fast photometer \citep{harding16} on 2020 November 13 and 14 (UT). Because the frame-transfer camera had no dead time between images, the cadence matched the 3~s exposure time. Images were obtained simultaneously through Sloan $g'$ and $i'$ filters. The length of the first time series was nearly an hour and the next night covered 1.6~hours. The images were bias subtracted and divided by twilight flat fields using the standard CHIMERA pipeline\footnote{\url{https://github.com/caltech-chimera/PyChimera}}. We used a variable aperture applied to the target and reference star of 1.5 times the seeing width to generate a differential light curve of the target.

\section{Analysis}

\subsection{Photometry}

\subsubsection{Eclipses of the flaring region \label{eclipse}}

We obtained $B$-band photometry of five eclipses predicted by the \citet{littlefield20} ephemeris and the light curves are shown in Figure~\ref{eclipse_lc}. As a guide to interpret the eclipses, we have generated a model eclipse light curve for a WD and a Roche lobe-filling secondary that assumes a mass ratio of $q=0.5$ \citep{eggleton83}. Eclipses in polars tend to have very sharp ingress and egress due to the small size of the WD. The eclipses in \lamost\ during flaring shown in Figure~\ref{eclipse_lc} have a gradual ingress lasting 5 to 10 minutes. To soften the transition, the model uses a Gaussian distribution for the light source assumed to be located at the position of the WD. We adjusted the Gaussian parameters and found that a width of 0.15$a$ (FWHM), or 15\%\ of the stellar separation approximated the ingress length. Finally we adjusted the orbital inclination to $i=81^\circ$ so that the model approximately matches the length of the eclipse. Certainly, other combinations of parameters could also approximate the observed eclipse profile, but the goal for this model is simply to use it as a consistent reference to compare the various photometric and spectroscopic observations.

The appearance of the eclipses in Figure~\ref{eclipse_lc} depends profoundly on whether there is flaring activity near the time of eclipse. Three of the eclipses occurred during periods of flaring, and the outline of the eclipse is well-defined, confirming the identification of eclipses by \citet{littlefield20}. Conversely, the eclipse is not evident in the two runs during which there was little or no flaring. Nevertheless, the brightness of the system at inferior conjunction was extremely consistent in all five light curves, with the average measured at $B=19.01\pm 0.02$ mag. Even during strong flares, the brightness at mid-eclipse is within a mmag of the mid-eclipse brightness in the runs without significant flaring, which suggests that the region producing the continuum flares is totally blocked by the secondary.

Some of the $B$-band light curves show a slight rise at the 0.1~mag level immediately after inferior conjunction. Further, the eclipse egress during flaring is more shallow than the ingress. Both of these characteristics imply that the intensity distribution of the continuum flaring region may be asymmetric, yet sufficiently compact to be nearly totally obscured by the secondary star.

The photometry obtained by \citet{thorstensen20} near inferior conjunction hints at some variability in the eclipse width when compared with the very consistent eclipse profile seen in our data. Any comparisons are complicated by the rapid brightness changes in the flares and differences in the central wavelengths between the two sets of observations.  

\begin{figure}[h]
    \centering
    \includegraphics[width=\columnwidth]{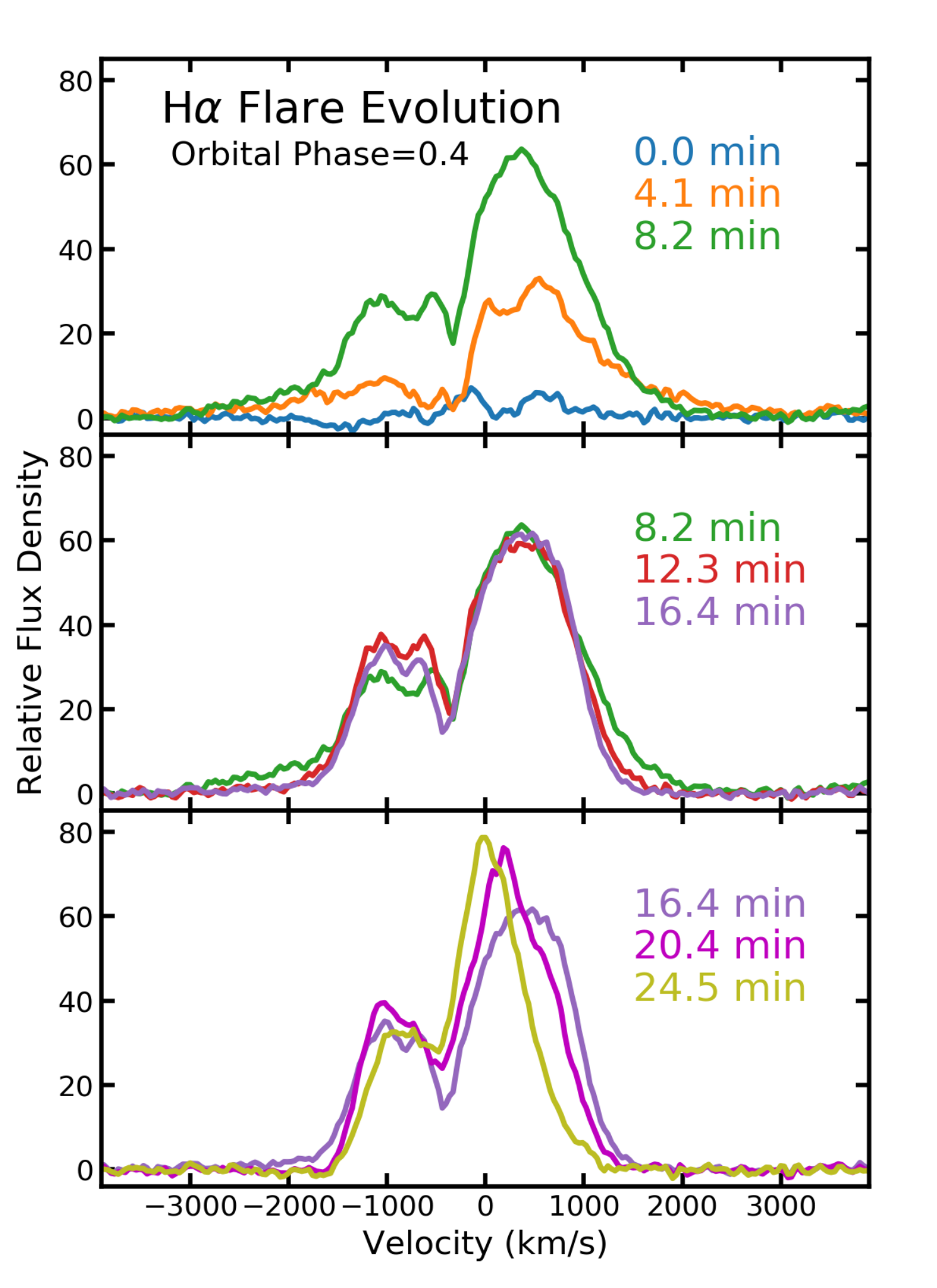}
    \caption{The evolution of the H$_\alpha$ emission profile during a September flare. The average spectrum just before the initiation of the flare was subtracted from each subsequent spectrum. The evolution is divided into three panels to show the changes clearly. The top panel displays the rapid rise of the flare and the development of the high-velocity wings. Red and blue-shifted emission is seen out to 3000~\kms\ during the rise, but the redshifted emission fades as the blue side continues to strengthen. Over the next 8 minutes (center panel), the H$_\alpha$ profile remains fairly constant except for the loss of the highest velocity wings. A narrow absorption is seen over the flare. Its blueshift increases from $300$~\kms\ to 400~\kms\ over 25 minutes. The lower panel shows the emission narrowing while the flux remains well above the pre-flare level.  }
    \label{evolution}
\end{figure}

\subsubsection{Non-detection of a spin signal} 

To search for the spin period of the WD, we obtained high-cadence photometry of \lamost, with the goal of detecting the spin modulation. \citet{thorstensen20} found no evidence of a spin period in his 23.3~s cadence photometry. For each night's time series, we computed the Lomb-Scargle power spectrum up to the Nyquist frequency of the sampling, but we also did not detect any significant candidate spin periods in the power spectra. Because the Palomar photometry was obtained on consecutive nights, we combined them into one dataset and performed the Lomb-Scargle analysis, but as before, there was no sign of coherent variability.

To estimate a limit on the amplitude of any spin signal, we removed the low-frequency variations in the P200 data and randomized the magnitude measurements before calculating a power spectrum. We repeated this process 500 times, recording the maximum power between frequencies 0.7$< f < 9.5$ cycles/min in each randomized light curve. Injecting a sinusoid signal with a 4~mmag amplitude resulted in a peak at the 3$\sigma$ level in the combined $g$-band P200 data. We conclude that any spin modulation with a period between 6.3~s and 85~s must have an amplitude below 4~mmag during our observation. 

Because AE~Aqr flares can display relatively high-amplitude quasi-periodic oscillations (QPOs) close to the spin period \citep{patterson79}, we also created two-dimensional power spectra of each light curve in order to search for intermittent periodic variations during the flares (Figure~\ref{fig:trailed_power}). However, we do not see any evidence of a QPO.

\subsection{Spectroscopy}

\subsubsection{Flares}

While collecting the data, it was obvious that the star was varying rapidly in brightness, color, emission line flux, and line width. The synthetic $R$-band magnitude variations shows the flaring pointed out by \citet{thorstensen20}, and it is consistent with that seen in AE~Aqr.

The H$_\alpha$ emission profile over a full binary orbit is shown in Figure~\ref{stack}, although the orbit was constructed from two runs separated by a month. Over the September run, the system slowly brightens from inferior conjunction (orbital phase $\phi =0.0$ based on the \citealt{littlefield20} ephemeris). Some flares are seen bracketing inferior conjunction as well as a major flare at $\phi =0.40$. The width of the H$_\alpha$ emission line narrows as the flare fades. The continuum reaches a deep minimum around superior conjunction ($\phi =0.5$) that is likely due to a lack of flaring combined with the ellipsoidal shape of the secondary. The September $LBT$ sequence transitions to the October run at $\phi =0.51$. A narrow P~Cygni-like (P~Cyg) absorption feature is seen through the October spectral series. Flaring in \lamost\ was more active in October than in September, displaying many overlapping events and a bright continuum.

In AE~Aqr, the highest emission velocities during flares reach $\pm 2000$ \kms \citep{welsh98}. In comparison, emission at the peak of the flares for \lamost\ extends to $\pm 3000$~\kms.  These high velocities last only a few spectra, or approximately 10~min. In isolated flares (see Figure~\ref{stack}), the continuum flux fades more quickly than the emission line flux, a behavior also seen in AE~Aqr \citep{welsh98}.

Balmer emission is visible throughout the orbit, even during periods of relatively low activity. During quiescent periods between flares, H$_\alpha$ is seen with a substantial velocity width of 1200~\kms\ HWZF. Around inferior conjunction the H$_\alpha$ emission width reaches its minimum value of 980~\kms\ HWZF. From this behavior we infer that some level of flaring activity is visible at all times, except when the flaring region is obscured by the secondary star at inferior conjunction.

\begin{figure*}
    \centering
    \includegraphics[width=\textwidth]{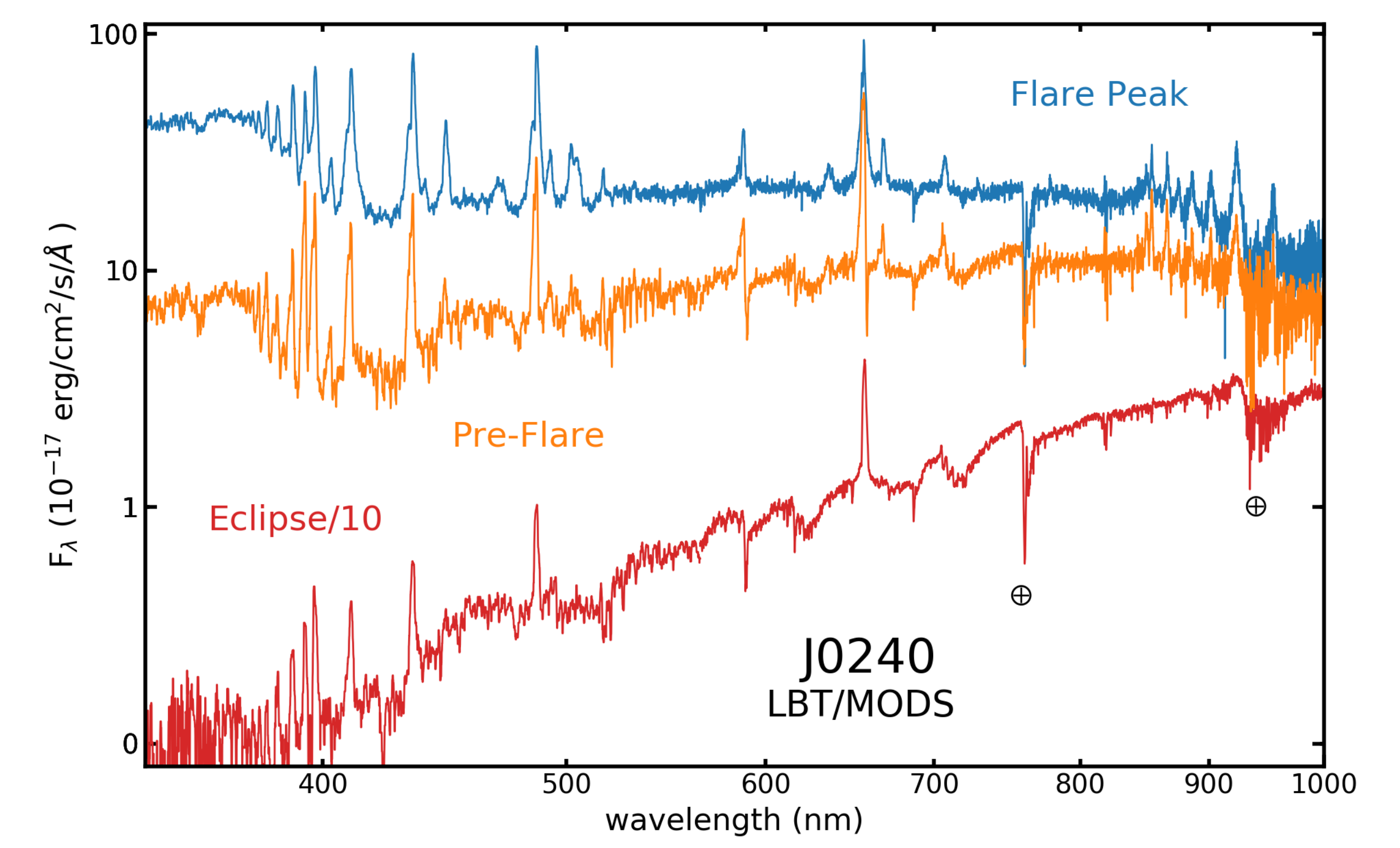}
    \caption{Example spectra of \lamost\ obtained during the inferior conjunction (eclipse; red), before a flare (orange), and at the peak of a flare (blue). The flare spectra are shown with the eclipse spectrum subtracted. The eclipse spectrum is dominated by the early M-type secondary, narrow Balmer and Ca~II emission (note the eclipse spectrum flux has been divided by 10). }
    \label{spec}
\end{figure*}

During the September sequence, a fairly isolated flare occurred that allows us to examine its evolution. We subtracted the H$_\alpha$ profile obtained immediately before the onset of the flare that began at orbital phase 0.40 and the results are displayed in Figure~\ref{evolution}. As noted above, the high-velocity wings are seen early in the flare and fade quickly, on the time-scale of the continuum decline. The emission profile is always asymmetric with the red side twice the flux of the blue. This asymmetry suggests that the blue-shifted emission is self-absorbed in an expanding wind. Some asymmetry in the line shape is seen in AE~Aqr \citep{welsh98}, but the high inclination of \lamost\ may increase this effect.

Once the flare reaches its peak emission flux the line shape remains fairly constant except for the loss of the high-velocity wings. An unresolved absorption feature is seen throughout the flare. Its blue-shifted velocity increases from 300 to 400~\kms\ over the 25~min. This absorption may correspond to the P-Cyg features seen during the second half of the orbit as its velocity is consistent with an extrapolation from the later orbital phases. 

In Figure~\ref{spec}, we show representative spectra of \lamost\ in three different intervals: at the maximum of a flare, before the flare, and during eclipse. Before a flare, Balmer, Paschen, He~I, and Ca~II emission lines are visible. During the flares, the emission lines broaden and intensify. The strengths of He~I lines are enhanced, weak He~II $\lambda4686$\AA\ emission briefly appears as well as the Si~II $\lambda6340$\AA\ doublet. The continuum becomes bluer during the flares with an increasing continuum across the Balmer jump into the ultraviolet (UV).
 
\subsubsection{Eclipse spectra}

Our $LBT$ spectra obtained during an eclipse offer additional insight into the eclipse photometry. The eclipse spectra  indicate that the H$_\alpha$ emission line flux does not match the profile of the continuum eclipse (Figure~\ref{eclipse_lc}). The line flux rises rapidly beginning at inferior conjunction, suggesting that the emission lines may be slightly offset from the location of continuum flares, or extend further from the flare location than the continuum emission. As seen in both AE~Aqr and \lamost\ flares, the emission lines fade more slowly than the continuum flux, implying that the gas producing the emission lines may flow out of the location of the continuum flare.

The eclipse spectrum is dominated by the secondary star (Figure~\ref{spec}), and the continuum is very red as expected for the early M-type companion \citep{thorstensen20}. Weak Balmer and Ca~II emission lines remain visible through eclipse, but the He~I lines are not detectable. This supports the proposal by \citet{littlefield20} that the main He~I-emitting region is blocked by the donor star. 

During the eclipse, the Balmer decrement becomes very steep in comparison to flare and pre-flare spectra. The Balmer decrement, $D_{3,4}=F(H_\alpha )/F(H_\beta)$, is 5.4$\pm 0.2$ in eclipse, $D_{3,4} =3.3\pm 0.1$ before a flare, and $D_{3,4} =2.1\pm 0.1$ during the peak of a flare. Since the eclipse blocks the source of the continuum flaring activity, it is likely that the Balmer emission visible during an eclipse comes from gas ejected by the propeller and is moving away from the binary. While interstellar dust reddening can be a cause of a steep Balmer decrement, for \lamost\ the reddening is very small \citep[see][]{thorstensen20}. We propose that the steep decrement results from a very large H$_\alpha$ optical depth \citep[e.g.][]{netzer75} caused by viewing the circumbinary gas close to the orbital plane. 




\subsubsection{Color Variations}

\begin{figure}[h!]
    \centering
    \includegraphics[width=\columnwidth]{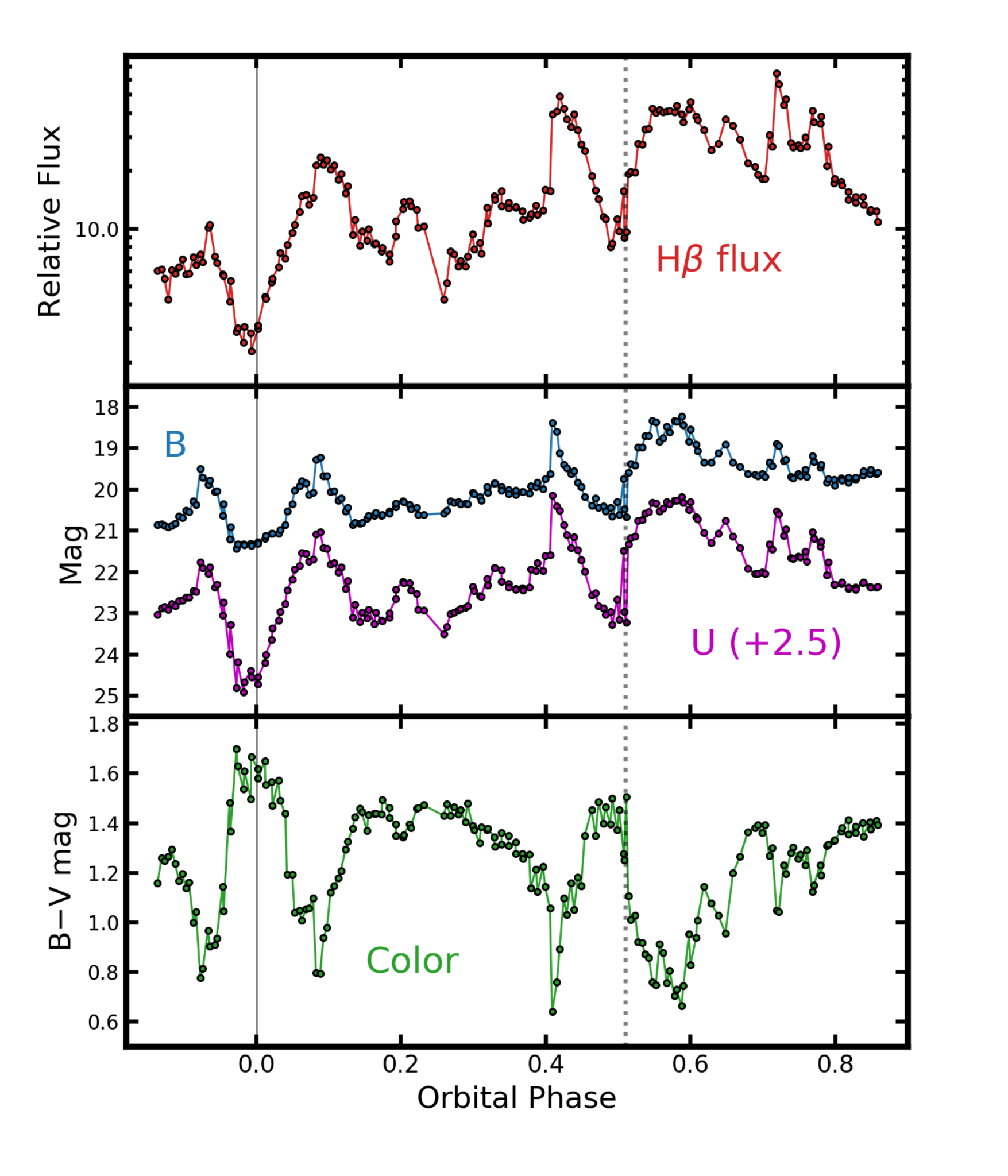}
    \caption{Light curves derived from the September and October spectroscopy. In all the panels the dotted vertical line indicates the transition between the two runs and the solid line is the time of inferior conjunction.  {\bf top:} Relative flux of H$_\beta$ versus orbital phase. {\bf middle:} Synthesized $U$ and $B$ magnitudes versus orbital phase. {\bf bottom:} Color index of the synthesized $B$ and $V$ magnitudes versus orbital phase. The color index reaches 0.8 to 0.6 mag during flares. Around eclipse the color is 1.6~mag, consistent with the M type secondary star.     }
    \label{blue_lc}
\end{figure}

The photometry synthesized from the spectra and compared with the variations in the emission line fluxes are shown in Figure~\ref{blue_lc}. Flaring is seen around inferior conjunction and at orbital phase 0.4, as well as nearly continuous activity in the October run that covered phases 0.50 to 0.85. 

The Balmer emission line flux and continuum brightness rise quickly, on time scales of a few minutes, then fade more slowly. The emission lines clearly take longer to fade after a flare when compared with the continuum flux. The $U$-band variations are more similar to the emission line changes than to the $B$-band fluctuations. For example, the H$_\beta$ flux and $U$-band brightness begin to rise at the time of inferior conjunction (orbital phase 0.0), while the $B$-band and $B-V$ color are relatively flat around conjunction.

The $B-V$ color index is reddest near inferior conjunction suggesting that the eclipse is blocking nearly all of the flaring activity. The color index at eclipse is $B-V = 1.6\pm 0.1$ mag, consistent with the color of an early M-type star \citep{thorstensen20}. During quiescent periods outside of eclipse the color index is 1.4 mag, while at the peak of flares the color reaches between 0.8 and 0.6~mag. Simultaneous multi-band photometry of AE~Aqr shows color variations between $1.0 > B-V > 0.5$ mag \citep{zamanov17}. The secondary in AE~Aqr is hotter than in \lamost , corresponding to an early K-type spectrum \citep{echevarria08}.  

\begin{figure}[h!]
    \centering
    \includegraphics[width=\columnwidth]{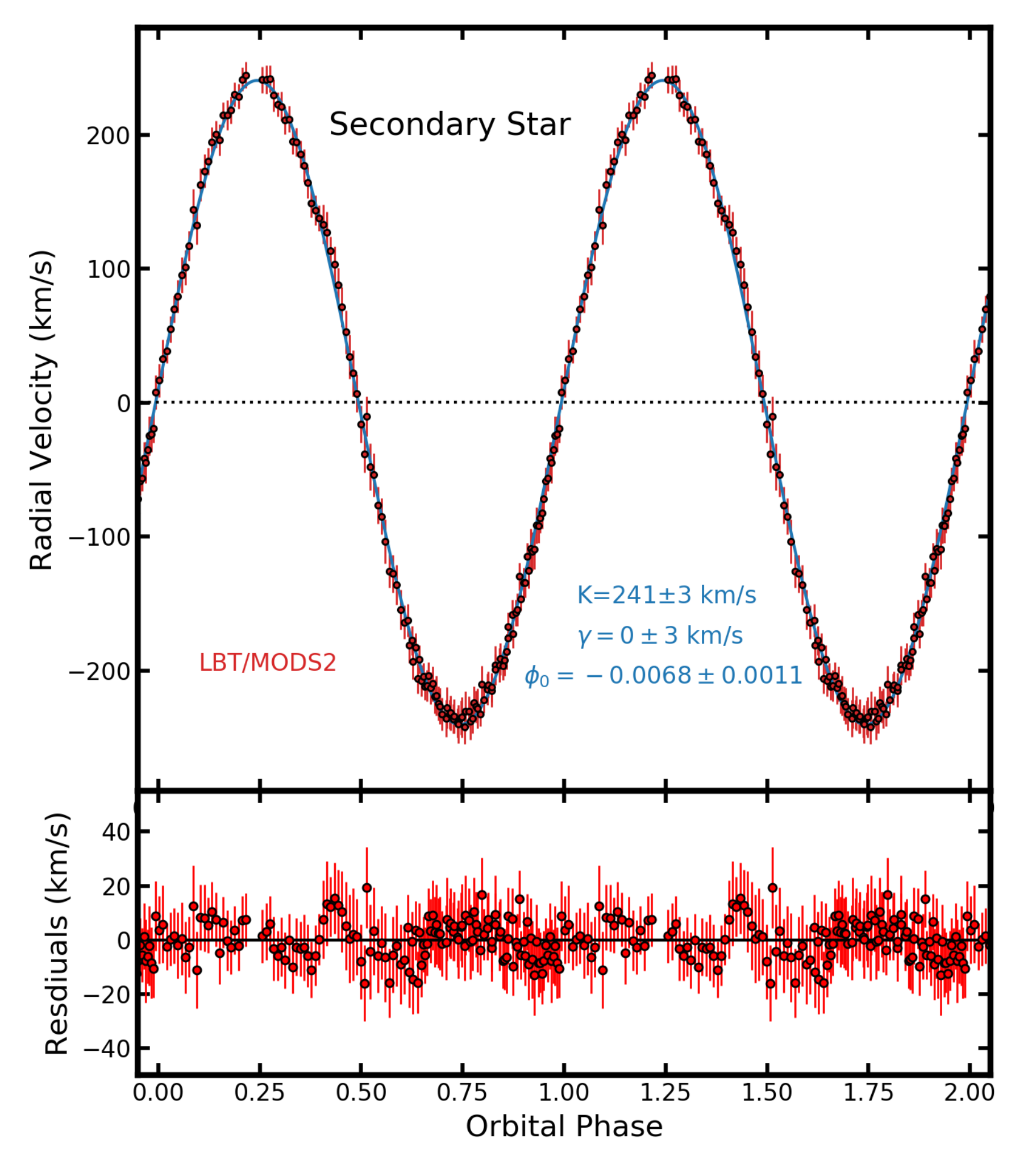}
    \caption{ {\bf Top:} The radial velocity estimates for the secondary star as a function of orbital phase. The points are measurements from individual MODS2 spectra obtained on three separate nights. The blue line is the best fit sinusoid assuming a fixed orbital period from \citet{littlefield20}. The data show that inferior conjunction occurs 3.0$\pm 0.5 $ minutes earlier than mid-eclipse predicted by the \citet{littlefield20} photometric ephemeris. {\bf Bottom:} Residuals between the data and the best fit sinusoid. The large flare seen at orbital phase 0.4 impacted the cross-correlation velocities and generated a 10~\kms\ amplitude wave in the radial velocity curve. }
    \label{secondary}
\end{figure}

\subsubsection{Orbital Motion of the Secondary}

Absorption features from the M-star secondary are clearly seem in all the $LBT$ spectra. This allows us to measure the orbital motion of the secondary star with the goal of precisely measuring the time of inferior conjunction relative to the eclipse ephemeris \citep{littlefield20}. \citet{thorstensen20} classified the secondary star as an M1.5 dwarf, so we cross-correlated sections of the $LBT$ spectra with a late stellar spectrum synthesized from the MILES stellar library \citep{vazdekis10}\footnote{http://research.iac.es/proyecto/miles/}.

Cross-correlations between the spectra and the template were performed using the {\bf fxcor} routine implemented in IRAF. Only the red arm of $MODS2$ was used for this analysis as it has the highest signal-to-noise ratio in the continuum of the four spectrographs. To avoid emission lines and reduce telluric contamination, only the wavelength ranges 6000-6530~\AA\ and 7800-8400~\AA\ were used in the cross-correlations. Over the three observing nights, 143 spectra were measured covering an entire orbital cycle with redundant measurements between phases 0.6 and 0.9. The typical uncertainty of each measurement was 11~\kms . A heliocentric correction was applied to each spectrum and the resulting radial velocity curve for the \lamost\ secondary star is shown in Figure~\ref{secondary}.

We converted the barycentric Julian day (BJD) at the center of each exposure to photometric phase using the \citet{littlefield20} ephemeris and fit a sinusoid function to the data using a least-squares algorithm. The half amplitude of the radial velocity curve is 241$\pm 3$~\kms , which is consistent with the \citet{thorstensen20} estimate of 250$\pm 8$~\kms . Inferior conjunction was found to occur at phase $\phi = -0.0068\pm 0.0011$ relative to the photometric ephemeris. This is very close to the predicted center of the eclipse, but 3.0$\pm 0.5$ minutes earlier than the photometric ephemeris time. The small size of this offset implies that the location of the flares is near the radius vector (the line passing through both stars and the system center of mass). Combining our measurement of $T_0$ with the \citet{littlefield20} orbital period, we obtain a new ephemeris of
\begin{equation}
T_{conj} = 2459105.84663(3) + 0.3056849(5)\times E, \label{ephemeris}
\end{equation}
where $T_{conj}$ is the time of inferior conjunction expressed as a Barycentric Julian Date in Barycentric Dynamical Time (TDB) and $E$ is the integer cycle count. The precise time of inferior conjunction defines the orbital phase of the secondary star and this is critical in narrowing the location of the flaring emission in the binary system.

The radial velocity curve shown in Figure~\ref{secondary} shows a small deformation at orbital phase 0.40. The amplitude of the wave is about 10~\kms . This was a phase covered in the 2020 September run and is coincident with a large flare seen in the continuum and in the emission lines. While the cross-correlation segments avoided major emission features, some lines such as the Si~II 6340~\AA\ are seen in emission only during flares. These weaker emission features appear to have a mild impact on the absorption line cross-correlation velocity estimates.

\begin{figure}[h!]
    \centering
    \includegraphics[width=\columnwidth]{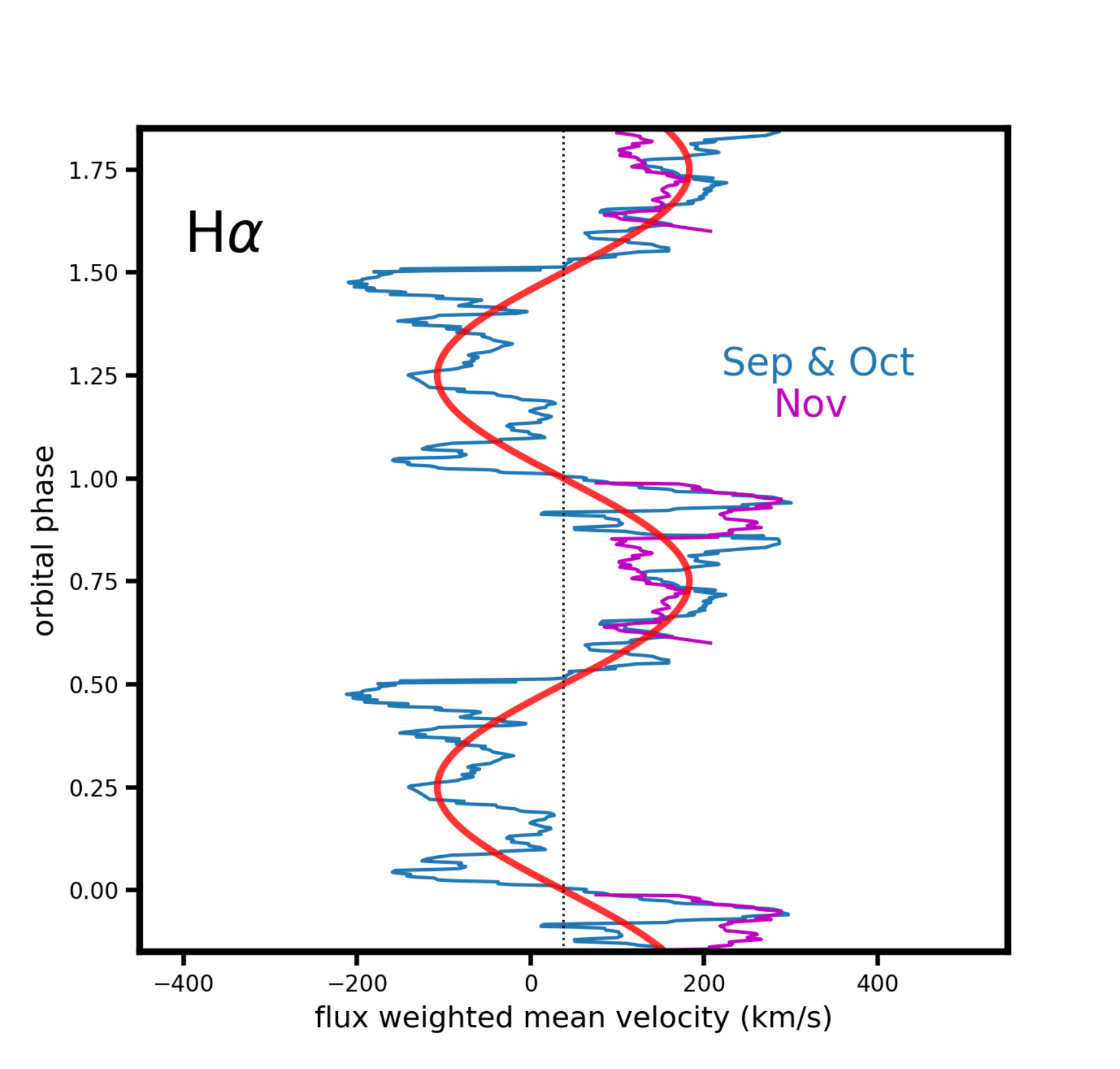}
    \caption{The flux-weighted average H$_\alpha$ velocity versus orbital phase (repeated for two orbits). The November run overlaps with the October data so it is shown in magenta. The red line is a best-fit sinuosoid with an amplitude of 145~\kms\ and the opposite phase of the secondary star. The zero point used in plotting the velocities is the systemic radial velocity, found from  the mean velocity of the secondary's absorption lines. }
    \label{ha_vel}
\end{figure}

\subsubsection{Orbital Motion of the Emission Line Region}\label{emission_vel}

A close inspection of the trailed H$_\alpha$ spectrum in Figure~\ref{stack} suggests a pattern in the velocity of the emission centroid with orbital phase. The emission appears most redshifted near $\phi \approx 0.75$ and mildly blueshifted around $\phi \approx 0.25$, although the high velocity flares and P~Cyg absorption makes this difficult to detect. In disk systems, a ``double Gaussian'' is often used to detect the orbital motion of the inner disk by focusing on the high-velocity wings of emission lines. For AE~Aqr and \lamost , the high velocity emission likely comes from shocked gas that would not be a good tracer of orbital motion, and application of this technique may have resulted in the conflicting phase estimates of \citet{robinson91} and \citet{welsh98}. Instead, we measure the flux-weighted mean wavelength of the H$_\alpha$ emission feature. The results are shown in Figure~\ref{ha_vel}. 

The shifting centroid of the H$_\alpha$ emission moves in anti-phase to that of the secondary star's motion. The velocity centroid is redshifted over the orbital phases $0.5 < \phi < 1.0$ while the secondary is blue-shifted over those phases. A sinusoidal fit to the centroid velocity suggests an offset of only 0.045 in orbital phase when compared to the secondary star. The center-of-mass appears shifted to the red by 40~\kms , but this shift may result from the asymmetry of the emission caused by the P~Cyg absorption and other optical depth effects that have a major impact on the blue side of the line.

We propose that the H$_\alpha$ line emitting region is nearly fixed in the binary frame and orbiting about the system's center of mass. As the H$_\alpha$ emission has about the same phasing as expected for the WD, the emission would be originating within 20$^\circ$ of the radius vector connecting the two stars and on the WD side of the system. Indeed, the velocity amplitude is close to what is expected for the WD given the orbital period of this system. To illustrate this point, from the velocity amplitude of the secondary, $V_2$, and its orbital angular velocity ($\Omega^{-1}$=4203.5 sec~rad$^{-1}$), we can estimate the physical distance of the secondary from the center-of-mass, $r_2$, as
\begin{equation}
r_2\; = V_2/(\Omega\; {\rm sin}(i))
\end{equation}
and for a mass ratio $q$, the WD separation from the center-of-mass is simply $r_1\; =q\; r_2$. The parameter of interest is the location  of the emitting region relative to the WD, or,
\begin{equation}
r_{Ha}/r_1\; = V_{Ha}/(qV_2)
\end{equation}
where $r_{Ha}/r_1$ is the ratio of the center-of-mass distances of the emitting region and WD and $V_{Ha}$ is the orbital amplitude of the emission region. The inclination dependence has cancelled, but it would be small anyway given that \lamost\ is eclipsing. The velocities have been estimated as $V_2=241$~\kms\ and $V_{Ha}=145$~\kms , so $r_{Ha}/r_1\;=0.6/q$. If the velocity amplitude of the H$_\alpha$ emission is a good representation of its orbital motion, then the emitting region is close to the WD for reasonable mass ratios. For example, for $q\approx 0.5$, the emitting region is 20\%\ further from the center-of-mass than the WD. From this analysis, it is unlikely that ejected blobs colliding at several stellar separations from the center of mass would produce such a low velocity amplitude and be in phase with the WD orbit.

%

\begin{figure}[h!]
    \centering
    \includegraphics[width=\columnwidth]{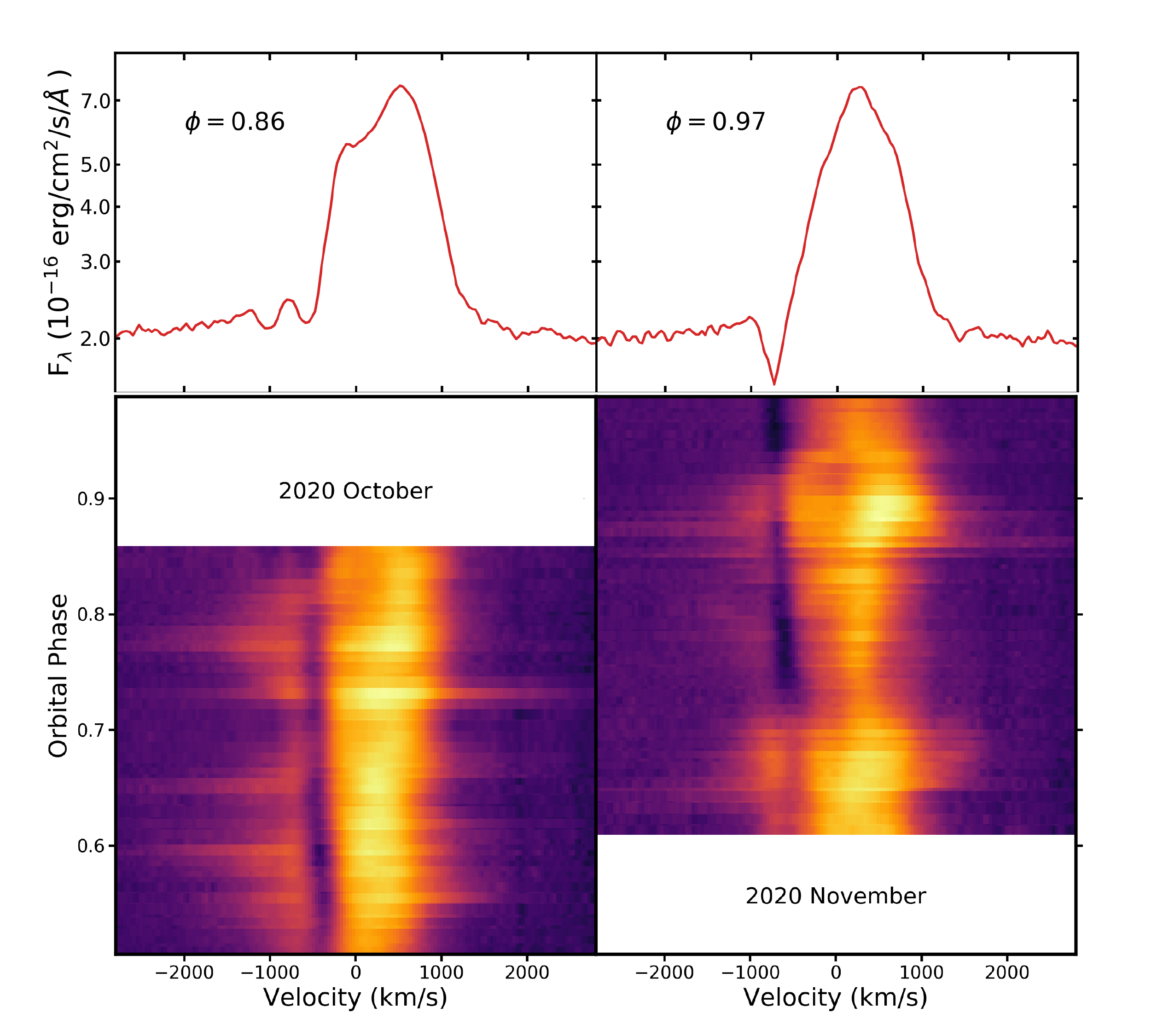}
    \caption{{\bf bottom:}  Trailed spectra centered on the H$_\alpha$ emission from $LBT$ data taken 2020 October (left) and 2020 November (right).  The narrow blueshifted absorption feature is seen continuously over the second half of the orbit. {\bf top:} Individual spectra from the two data sets. On the left is a spectrum showing a pair of narrow absorption lines at $-550$ and $-1025$~\kms . Displayed on the right is a spectrum obtained near inferior conjunction where the absorption is well below the continuum of the secondary star. The deep absorption at this orbital phase suggests that the intervening gas is on the opposite side of the system from the WD and in circumbinary orbit after acceleration by the propeller.  }
    \label{composite}
\end{figure}

\subsubsection{Blue-shifted Absorption Feature}

\citet{thorstensen20} noted a strong, blueshifted absorption feature that he associated with periods of flaring. Our spectra reveal that this absorption is present---even when \lamost\ is not flaring---across a wide range of orbital phases (Figure~\ref{composite}), particularly between orbital phases 0.5$<\phi <$1.0. The blueshifted absorption feature provides direct evidence of an outflow, and as Sec.~\ref{sec:model} will explain in detail using the \citet{wynn97} modeling, we conclude that it occurs in J0240 when the gas expelled by the magnetic propeller passes in front of the secondary and/or the flaring region. Similar absorption features have not been observed in AE~Aqr, due to its much lower inclination. Because the absorption directly probes the conditions in the outflow, it is especially important to characterize it properties.

\begin{figure}[h!]
    \centering
    \includegraphics[width=\columnwidth]{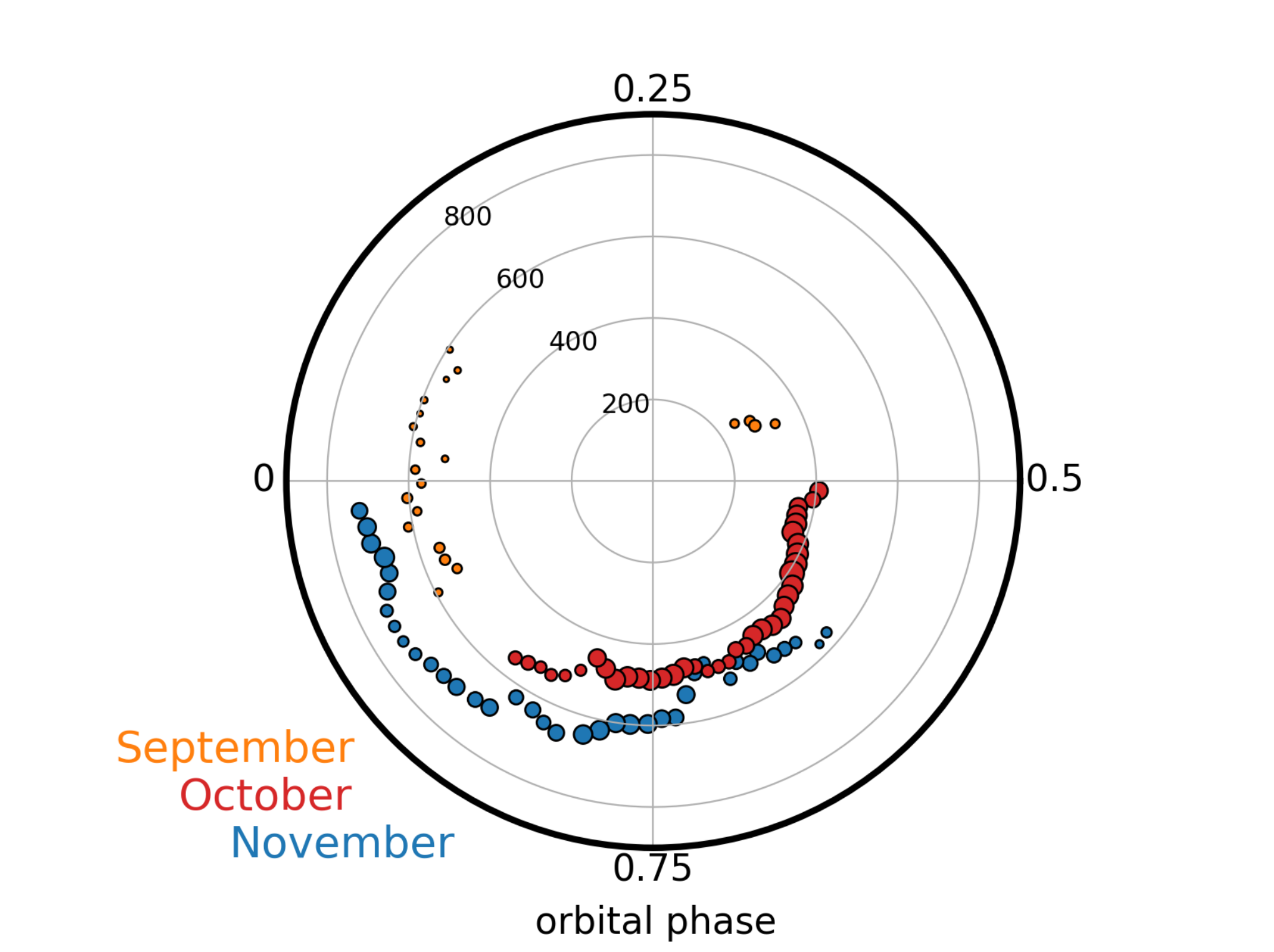}
    \caption{The blue-shifted velocity (in \kms ) of the narrow P~Cyg feature as a function of orbital phase depicted on a polar projection. The spectra obtained in November (blue points) shows a slightly larger velocity over the same orbital phases when compared with the September (orange) and October (red) data. The size of each point is proportional to the equivalent width of the absorption. The P~Cyg absorption feature is most clearly seen over the second half of the orbit. 
    }
    \label{pcyg}
\end{figure}

The most obvious feature of the blueshifted absorption is that it is present during the second half of the orbit, as is particularly apparent in the October and November spectra (Figure~\ref{composite}). During that time, the absorption varies slowly in velocity, and in general, the absorption minima are seen to become more negative with orbital phase, starting around $-380$~\kms\ at $\phi\sim 0.5$ and reaching to $-600$ or $-700$~\kms\ approaching inferior conjunction. In Figure~\ref{pcyg}, we plot the blueshift of the absorption (when it was present) against the orbital phase of the observation. During the November run, the minimum of the absorption is seen at slightly higher velocities than measured in the October and September data, suggesting that its properties vary over time. 

A few spectra showed multiple blue-shifted absorption components, with the velocity of the additional components between 1000 to 1200~\kms . An example of a double absorption feature is shown in Figure~\ref{composite} at $\phi =0.86$. The high-velocity component was detectable for approximately 15~minutes.

The equivalent widths (EW) of the absorption can be as high as 4~\AA , but typically the EW are between 1 and 3~\AA\ over the second half of the orbit. When the system is not flaring, the absorption falls below the continuum of the secondary. This is clearly seen in the spectrum taken near inferior conjunction and shown in Figure~\ref{composite}.  Because the flaring region is completely hidden during the eclipse (see Sec.~\ref{eclipse}), the fact that the P~Cyg absorption cuts into the donor's continuum demonstrates that the absorption must originate in gas on the opposite side of the secondary from the WD.

The width of the absorption feature is difficult to measure precisely because of the steep continuum slope on the line wings. The absorption full-width at half minimum (FWHM) corresponds to 230$\pm 30$~\kms , which is similar to the FWHM of the night sky lines, implying that the absorption feature is unresolved. 

\section{Discussion}

\subsection{Location of the Flares}

The presence of eclipses in \lamost\ provides an opportunity to estimate the location of the flaring activity in the binary system. One of the longstanding debates with AE~Aqr is whether its flares are produced when blobs are shocked in the WD's magnetosphere \citep{eracleous96} or when they collide in the propeller's outflow, well past the WD \citep{welsh98}. The major difference between these two scenarios is that in the latter, the flares are produced at a large distance from the WD. 

We confirm the inference by \citet{littlefield20} that there is no visible flaring in optical broadband photometry during the eclipses. Furthermore, our spectra verify that mid-eclipse coincides with the secondary's inferior conjunction, implying that the flares are constrained to occur near the line passing through the two stars. This strongly favors the \citet{eracleous96} model, in which the flares are predicted to be generated in the WD magnetosphere.

As pointed out in section~\ref{emission_vel}, the apparent orbital velocity amplitude of the H$_\alpha$ emission region is similar to that expected for the WD.  In addition, the radial velocity of the line emitting region is anti-phased to that of the secondary. Combining these observations suggests that the emission line emitting region is near the WD.

In the \citet{welsh98} blob-blob collision model, the expected flaring location would result in eclipses well before inferior conjunction. The large extent of the predicted interaction region would generate long partial eclipses that would be difficult to identify given the flaring variability. Furthermore, in the \citet{welsh98} scenario, the location of the collision region should vary because blobs follow different trajectories, depending on their size and density. We would therefore expect eclipses to vary in phase, rather than producing the sharply defined eclipse seen in survey photometry \citep{littlefield20}.

That said, the emission line and ultraviolet (UV) light curves generated from our spectra  (see Figure~\ref{blue_lc}), differ from the broad-band photometric eclipses shown in Figure~\ref{eclipse_lc}. The UV and line emission start to rise at zero orbital phase, while the continuum flux is relatively flat on either side of inferior conjunction.  The flat-bottomed continuum eclipses suggest that most of the continuum-emitting region is blocked by the secondary. The rise of H$_\alpha$  emission and UV flux during the continuum eclipse implies that the line-emitting region may be more extended and peeking around the secondary at the moment of inferior conjunction. 

This geometry is consistent with the \citet{eracleous96} model where blobs traveling ballistically from the secondary are shocked when encountering the WD magnetosphere. This occurs near the WD and the shocks generate the continuum flares. The accelerated blobs continue past the WD while expanding and cooling, resulting in Balmer emission on the leading side of the WD. This process is likely to be messy judging by the range of observed absorption velocities, but with the data we have in hand, the continuum flares appear to be confined to within $0.2a$ of the WD, where $a$ is binary separation. 

\subsection{Flare High Velocity Emission}

We interpret the extreme velocity dispersion observed during the flares (HWZI$\sim3000$~\kms) as the expansion velocity of the blobs when they are shocked, as opposed to the initial velocity of the blobs immediately following their acceleration by the propeller. We base this inference on the presence of the extreme, nearly-simultaneous redshifts and blueshifts observed around inferior conjunction, when the trajectory of the expelled blobs is roughly perpendicular to our line of sight. If we treat the blobs as simple particles following trajectories out of the system, their velocity vectors at certain orbital phases would be perpendicular to our line of sight, so we would observe high-velocity flares only at certain orbital phases. Conversely, if the blobs rapidly expand after being shocked, the resulting Doppler shifts would be observable even if the blobs were moving in the transverse direction. \citet{eracleous96} argued that shocked blobs could expand at several thousand \kms\ because they would be opaque to the X-rays produced internally by the shock-heated plasma.

The blueshifted absorption component provides a second argument against interpreting the HWZI as the initial velocity of the ejecta. The hyperbolic excess velocity $v_{term}$ of an object is given by $v_{term} = \sqrt{v^2_0 - v^2_{escape}}$, and we find that if $v_0\sim3000$~\kms, the outflow will decelerate only slightly. Since the outflow expands in the radial direction (see Sec.~\ref{sec:model}), the projected velocity of the outflow's P~Cyg absorption component would therefore be much larger than what is observed.

\subsection{The Lack of Coherent Oscillations}

In most characteristics, \lamost\ and AE~Aqr are extremely similar. A 33~s oscillation is seen in AE~Aqr associated with the spin period of its WD; however, our fast photometry does not detect a coherent oscillation from \lamost. For AE~Aqr, \citet{patterson79} found the spin and its harmonic with amplitudes ranging between 1 and 10~mmag, but generally the signals were 2 to 3~mmag, just below our limit of detection. Observations by \citet{bruch94} found that the spin oscillations in AE~Aqr fell below detectability for a month, suggesting that we might simply be unlucky in the timing for our search. 

From UV observations, \citet{eracleous94} modeled the spin modulations from AE~Aqr as hotspots on its WD. Their best fit placed the spots at mid-latitudes on the star in order to match the very large amplitude variations seen in the UV. In \lamost, any hotspots may be close to the spin axis, resulting in a smaller amplitude. It may even be that \lamost\ is a more efficient propeller than AE~Aqr, and no accretion heating is occurring at the magnetic poles. 

More sensitive searches for the WD spin signature are needed. 
In AE~Aqr, the amplitude of the spin modulation increases toward shorter wavelengths, so fast cadence photometry in the UV might be the best way to detect the WD spin period in \lamost .  

\begin{figure*}
    \centering
    \includegraphics[width=\textwidth]{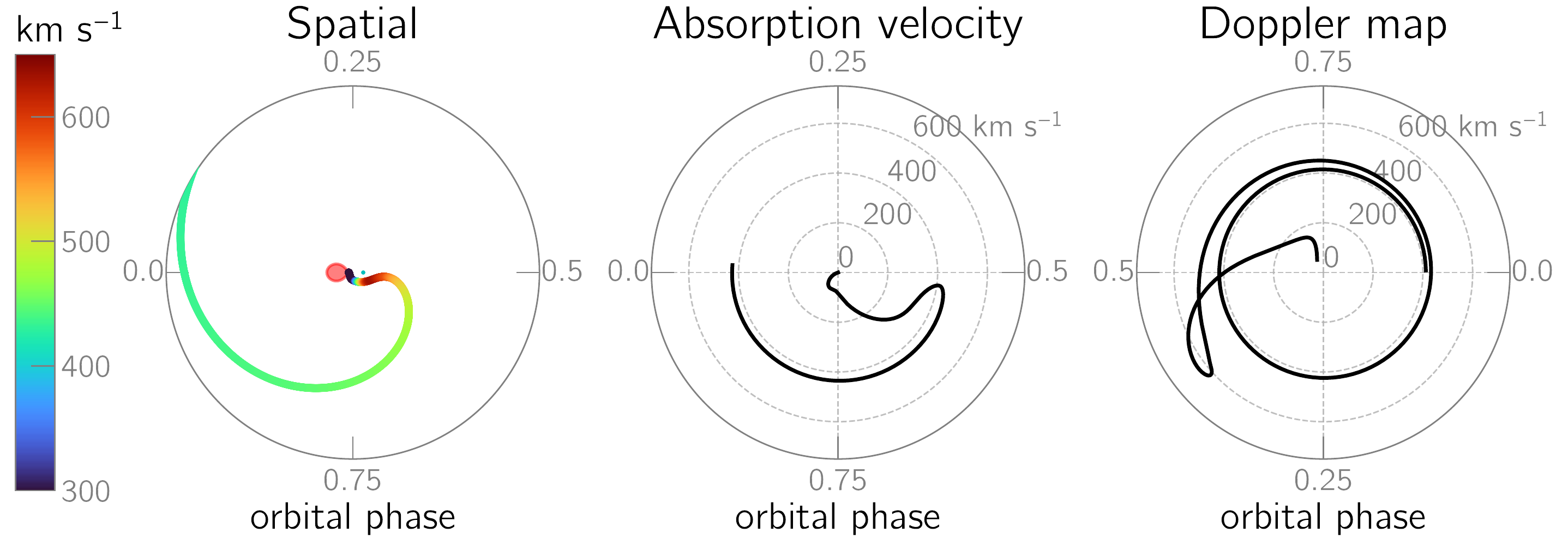}
    \caption{ {\bf Left:} Schematic diagram of an outflow that produces line absorption at high inclination, presented in the binary rest frame. The magnitude of the velocity vector is represented in color, and the numbers around the edge of the figure indicate orbital phases. The parameters used to create this model are discussed in the text.  {\bf Center:} The predicted radial velocity of an absorption feature resulting from the outflow model displayed in the left panel. This can be directly compared to the observed absorption velocities in \lamost\ displayed in Figure~\ref{pcyg}.  {\bf Right:} The motion of the outflow in velocity space. This panel describes emission from the outflow (not absorption), so the requirement of a backlight from the center panel is lifted.    }
    \label{fig:diagram}
\end{figure*}

\subsection{Application of \citet{wynn97} to \lamost} \label{sec:model}

Our results are strongly consistent with the model of AE Aqr from \citet{wynn97}, who argued that the accretion flow is clumpy, with discrete, heterogeneous blobs that each experience a drag force as they encounter the WD's magnetosphere. Near their point of closest approach to the WD, the blobs are accelerated above the binary escape velocity and expelled. As they leave the system, the blobs trace a spiral pattern in the binary rest frame. We refer to the blob velocity after ejection as the ``terminal'' velocity, but this is just an approximation as the blobs continue to slowly decelerate in the potential well of the binary.

Although the modeling in \citet{wynn97} is tailored to AE~Aqr specifically, nearly all of our observations can be understood within its framework. For example, in the \citet{wynn97} model, there is no accretion disk, and we see no evidence of one in our observations. Even more importantly, the persistent P~Cyg profile seen during the second half of the orbit is a natural consequence of the outflow trajectories computed by \citet{wynn97}. Their Figure~7 predicts that shortly after its interaction with the magnetosphere, the propeller's outflow will begin to spiral outward from the binary (as seen in the binary rest frame), and we would expect to see line absorption when this outflow lines up with our view of the secondary and flaring region. The velocity vectors in their Figure~7 are pointed outward radially, so the resulting absorption would be blueshifted by several hundred \kms, as observed. Unless the orbital inclination is exactly 90$^\circ$, the outflow will eventually be too far from the binary to intersect our line of sight to the secondary, and the absorption will therefore cease.

At present, \lamost\ cannot be modeled in the same way as AE~Aqr because its spin period is unknown. Therefore, to provide a quantitative basis for our interpretation, we shall consider how AE~Aqr would appear if it were viewed at a high orbital inclination. Blobs are expected to expand out of the orbital plane, and they may become comparable in size to the WD's Roche lobe after being expelled \citep{eracleous96}. In this scenario, ejected blobs could intercept the line between the observer and a source of light in the system, such as the secondary or flaring region. For example, at orbital phase 0.75, the propeller's outflow would be $\sim4a$ from the secondary in the modelling by \citet{wynn97}. If the secondary and one of the expelled blobs have radii of $0.3a$ and $0.4a$, respectively, the blob could occlude the secondary if $i \geq \tan^{-1}(4a/(0.3a+0.4a)) = 80^{\circ}$. A quarter-orbit later, when the outflow is $\sim5a$ from the secondary, the inclination required for an occlusion increases to $i \geq 82^{\circ}$. These simple geometric considerations suggest that $i$ dictates the range of orbital phases that the P~Cyg absorption component remains visible, because as the outflow becomes increasingly distant from the secondary, an increasingly high value of $i$ would be needed to produce an absorption feature.

Additionally, we note that a blob's path and its terminal velocity, can be tweaked by adjusting the drag coefficient, $k$, from \citet{wynn97}. Increasing this value causes the accretion stream to begin interacting further from the WD and results in a higher terminal velocity. We illustrate this for AE~Aqr using the parameters from \citet{wynn97} in Figure~\ref{fig:diagram}, which shows a single-particle trajectory for AE~Aqr for $k=2\times10^{-5}$~s$^{-1}$. Of particular note is the center panel of Figure~\ref{fig:diagram}, which shows the predicted radial velocity of a blob backlit by an emitting source at the binary center of mass. The center of mass was chosen for this calculation because it is conveniently situated between the two likely sources of light in the system: the secondary star and the flaring region near the WD. The predictions of the absorption velocities from gas close to the interaction region are unreliable in this approximation, but the absorption velocities from distant blobs do not strongly depend on the location of the light source.  

Figure~\ref{fig:diagram} displays a model for AE~Aqr, but its overall similarity to the P~Cyg velocities and orbital phases observed in \lamost\ (Figure~\ref{pcyg}) is quite striking.  There are certainly differences in the details between the observed and modelled absorption features. The terminal velocities in the AE~Aqr model are nearly constant between phases $0.5 < \phi < 1.0$ at about 400~\kms . In contrast, the observed \lamost\ absorption velocities are seen to increase from 400~\kms\ near $\phi =0.5$, to 600 to 700~\kms\ at inferior conjunction.  If the spin period were known, the $k$ parameter could be adjusted to achieve a better fit with the observed velocities. Nevertheless, with the parameters of AE~Aqr, we cannot find a value of $k$ that causes the predicted absorption velocity to increase at later orbital phases; indeed, this gas is far from the influence of the propeller and conservation of energy suggests that the expelled blobs would slowly decelerate as they move outward through the binary's gravitational potential. 

We speculate that the drag coefficient in a real propeller system may systematically vary on a time scale of minutes, hours, and days. This could result from variations in the blob properties as they leave the secondary, or from active flares modifying blobs approaching the WD.  We see kinks with amplitudes of 50~\kms\ in the absorption line velocities (Figure~\ref{pcyg}) over a single observing run, as well as differences of $\sim$100~\kms\ between runs, suggesting that the ejection velocities are varying at both short and long time scales. Further high-quality \lamost\ spectroscopy is needed to determine if the absorption velocities consistently increase with orbital phase as seen in two of our observing runs.

The ease of detecting the outflow in \lamost\ offers a stark but refreshing contrast with AE~Aqr, whose outflow is challenging to detect at that system's much lower inclination. Many previous studies have attempted to use Doppler tomography to discern the outflow \citep[e.g.][]{wynn97,welsh98,horne99}, but the resulting tomograms have tended to show little more than a featureless blob of emission centered in the $-V_{x}, -V_{y}$ quadrant. In contrast, the P~Cyg absorption in \lamost\ offers a direct and unambiguous means of studying the outflow. In particular, if a future study can identify the spin period as well as the orbital inclination $i$ from eclipse modeling, the exact trajectory of the outflow could be rigorously mapped. Likewise, theoretically predicted outflow velocities, such as those in Fig.~\ref{fig:diagram}, can be tested against the radial velocity of the P~Cyg absorption (see Figure~\ref{pcyg}) once the spin period is known.

\section{Conclusion}

Our observations and analysis establish that \lamost\ is a magnetic CV in a propeller state and the first eclipsing AE~Aqr type star. Our major conclusions are as follows.

\begin{enumerate}
    \item We confirm that the flaring region undergoes eclipses by the secondary star in \lamost\  as noted by \citet{littlefield20}.
    
    \item The optical flares noted by \citet{thorstensen20} coincide with transient emission-line flares whose wings extend to $\pm3000$~\kms\ in the Balmer and He~I lines. This unique high-velocity flaring is seen in AE~Aqr and points to a strong similarity between the two systems.
    
    \item We identify a persistent narrow Balmer absorption feature between orbital phases 0.5-1.0. Its blue-shifted velocity is seen to increase with orbital phase. We argue that this P~Cyg absorption results from gas ejected by the propeller into circumbinary orbit. 
    
    \item  The narrow absorption is seen below the level of the continuum from the secondary around inferior conjunction, showing that the absorbing gas is consistent with the outflow models of \citet{wynn97} and our own simulations.
    
    \item The emission lines are formed primarily in two distinct regions. One of them, which we identify as the magnetic-propeller region close to the WD, is the source of the higher-order Balmer, He~I, and high-velocity H$\alpha$ and H$\beta$ emission. The other is an extended outflow that produces mostly low-velocity H$\alpha$ emission, and its blue edge is defined with the presence of  P~Cyg absorption.
    
    \item We unsuccessfully searched for a photometric signature of the spinning WD as is seen in AE~Aqr. We placed a 4~mmag upper limit on the $g$-band spin amplitude for periods between 6.3~s and 85~s. This is consistent with the non-detection of a spin signal noted by \citet{pretorius21}.

\end{enumerate}

\lamost\ appears to be only the second member of the AE~Aqr sub-class of IPs. The proposed WD pulsar, AR~Sco \citep{marsh16,stiller18,garnavich19} may also be related to these AE~Aqr-like propellers. The propellers and AR~Sco appear to efficiently extract spin energy from their WDs, and are unusually strong radio emitters, implying a significant non-thermal power source derived from an interaction with the rapidly spinning magnetic WD. \citep{bookbinder87,stanway18,pretorius21}. 

Because of its high orbital inclination, \lamost\ enables observational tests that are impossible with AE~Aqr, which has a much lower orbital inclination. It is therefore a compelling candidate for theoretical modeling.

\begin{acknowledgements}

We thank R. Pogge and O. Kuhn for their help in obtaining the LBT observations.
This work is partly based on observations obtained at the MDM Observatory, operated by Dartmouth College, Columbia University, Ohio State University, Ohio University, and the University of Michigan.

\end{acknowledgements}
    

\end{document}